# Ion-liquid based super-capacitors with inner gate diode-like separators

Tazima S. Chowdhury and Haim Grebel*

*Electrical and Computer Engineering and Electronic Imaging Center, New Jersey Institute of Technology, Newark, NJ 07102, USA.*

**Abstract:**

We demonstrate that the capacitance of ionic-liquid filled supercapacitors is substantially increased by placing a diode-like structure on the separator membrane. We call the structured separator – gate, and demonstrate that the order of a p-n layout with respect to the auxiliary electrode affects the overall cell's capacitance. The smallest ESR and the largest capacitance values are noted when the p-side is facing the auxiliary electrode.

*\* Corresponding author*

*E-mail address*: grebel@njit.edu

## I. Introduction:

Increasingly, super-capacitors find more and more applications as short-term energy storage [1-5]. As for ordinary capacitors, increasing the surface area of the electrodes, decreasing the effective charge separation and in some cases, increasing the efficiency of charge transfer result in an increased overall cell capacitance. A separator between the electrodes (Fig. 1a) is introduced to minimize unintentional electrical discharge. We concentrate on the separator layer and modify the otherwise electrically insulating membrane [6-8] into a diode-like layer [9-11] (see Fig. 1b). The concept may be further developed into a transistor-like structure, as well [12]. Nanometer scale charge separation, charge transfer and electrodes porosity have been under intensive research [13-15]. Here we are concentrating on the separator membrane; the concept is general and may be easily incorporated into existing supercapacitor structures. Our diode-like separators are made of two layers, each composed of functionalized single-walled carbon nanotubes (SWCNT) film: one layer is composed of electronically p-type tubes and the other layer is composed of electronically n-type tubes. When pressed together, the interface between the layers behaves as a diode-like structure due to the numerous junctions that are formed between the two types [16]. The diode-like structure is permeable to ions and portrays resistance and capacitance.

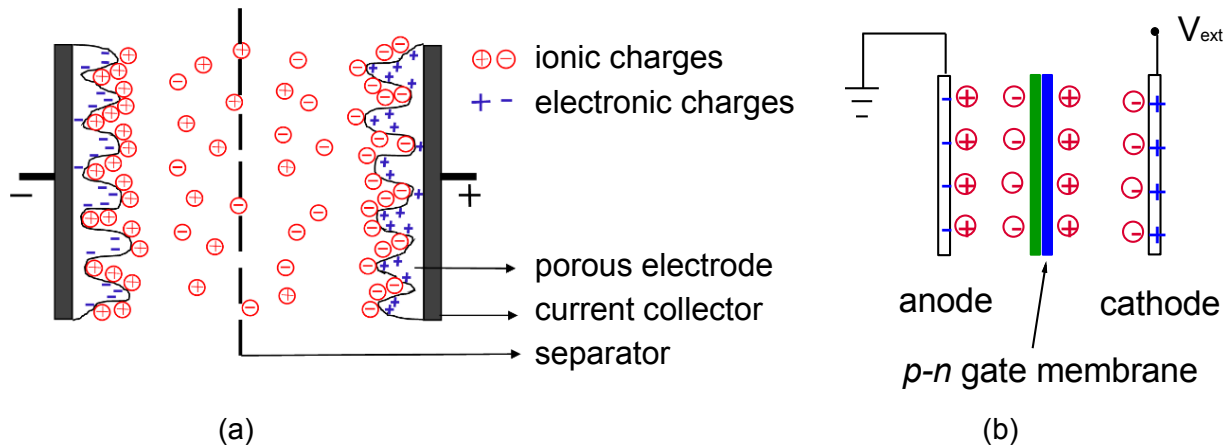

Fig. 1. (a) Schematics of super-capacitors. (b) Diode-like mid-cell gate structure: blue – electronic charges; red – ionic charges; green/blue layers – p-type/n-type layers.

## II. Experiment and Methods:

Single wall carbon nanotubes (SWCNT) were obtained from Nano Integris, Canada, with a purity better than 95%. The SWCNTs may be purchased with little content of metallic catalysts (<1%). The SWCNT were functionalized with polymers: *p*-type and *n*-type tubes were obtained by wrapping the tubes with PVP and PEI, respectively [17-18]. Functionalized single wall carbon nanotubes (SWCNT) have been used for the structured p-n gate element (Fig. 2) [19-22]. As shown in the figure, some of the SWCNT formed bundles which did not affect their electrical properties under dry conditions (see Fig. 3a-b). The p-type and n-type nanotubes were suspended in DI water using a horn probe sonicator for 8 hours. Using vacuum, each layer was drop-casted on a hydrophilic filter (TS80, TriSep). The thickness of each SWCNT film was estimated as a few microns by weight. Due to the relatively small bandgap at the p-n interface [16] and the relatively lightly doping of the tubes, the depletion region is expected to extend throughout the entire gate's cross-section.

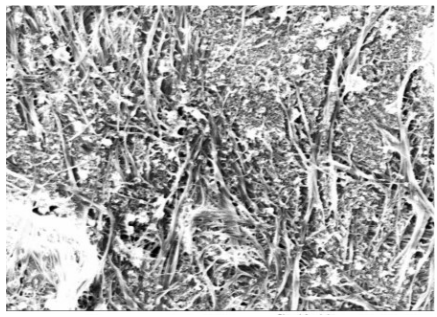

—— 1 $\mu$m

Fig. 2. SEM picture of a CNT film on top of a TS80 separator membrane.

Current-voltage (I-V) measurements were conducted on each SWCNT-type film (Fig. 3). A single type film behaves as a resistor-like, exhibiting a linear curve. The I-V curve for a diode-like structure behaves nonlinearly with an accelerated behavior in a forward direction (namely, when the positive lead is on the p-type film and the negative lead on the n-type side.

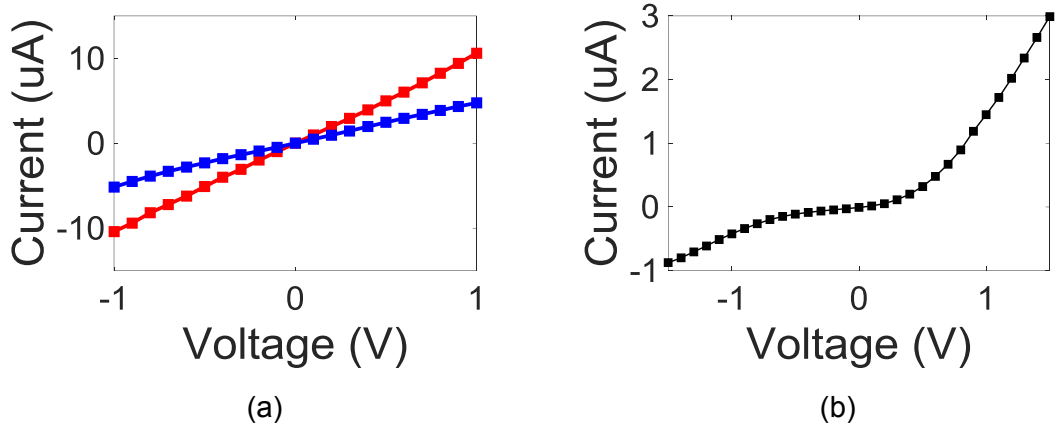

(a) (b)

Fig. 3. Electronic behavior. (a) Current-voltage (I-V) curve for a dry single CNT film on TS80: p-type (red) and n-type (blue). (b) I-V curve for a pressed p-n junction(s). In obtaining this curve, one contact was placed on the p-type side and the other contact was placed on the n-type side.

In Fig. 4 we show Raman spectra for our films. Both films exhibit a similar $G^+$ line at 1616 cm$^{-1}$. The broadened $G^-$ line is shifted due to the tube's doping: the n-type film exhibits a downshift of ca 10 cm$^{-1}$ due to its negative doping with respect to the p-type film. 3 mW HeNe laser at 633 nm was used in conjunction with a 75 cm spectrometer and a CCD array, cooled to -35 °C. The spectral resolution was 1 cm$^{-1}$.

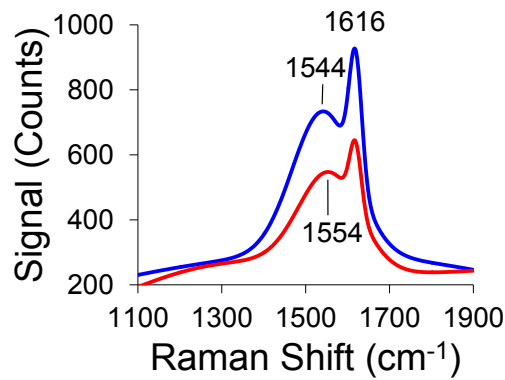

Fig. 4. Raman spectrum for single-type SWCNT films: n-type (blue) and p-type (red).

Thermoelectric data [16] are provided in Table 1 for each single-type film. Shown is the potential difference between a heated end (using a hot plate) and a cold end (at room temperature). The positive lead of the multi-meter was placed on the hot end. The data corroborate the doping type of the SWCNT; namely, negative values for a p-type film and positive values for an n-type film.

| Type | $\Delta V$ at 57 °C | $\Delta V$ at 68 °C |
|---|---|---|
| n-type | +0.8mV | +1.2mV |
| p-type | -1.8mV | -2.5mV |

Table 1. Thermoelectric measurements for single-type SWCNT films.

Ionic liquid, 1-n-Butyl-3-methyl-imidazolium hexafluro-phosphate, was used as an electrolyte. It filled the space between two 25-micron thick flat copper (Cu) electrodes of size 5x3 cm$^2$ each. The ionic liquid was soaking 0.1 mm thick lens tissues (Bausch & Lomb) used as spacers. In retrospect, one may eliminate the spacers and soak the TS80 membranes since the single-type SWCNT are coating only one side of the membrane. The configurations for the cell with and without the gate is shown in Fig. 5a-b. The assembled sample was let resting overnight. Modified Cu electrodes (Fig. 5c) were prepared by placing a double sided carbon tape (typically used for SEM samples) on the copper films and spreading 1-micron carbon powder (Alpha) on the tape. Loose powder was scraped off. These electrodes were not meant to be optimized highly porous electrodes; they helped us achieve reasonable capacitance values.

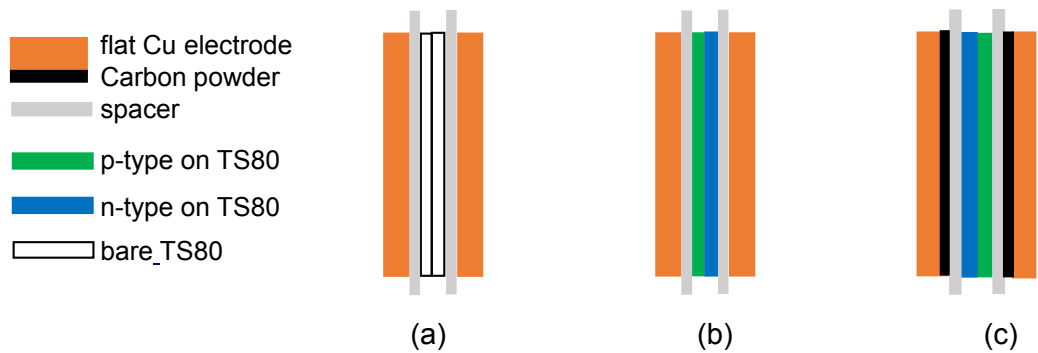

Fig. 5. Cell configuration with (a) bare, (b) structured separator (gate) with plain Cu electrodes and (c) structured separator with modified Cu electrodes.

### III. Results and discussion:

2-electrode CV traces and their corresponding capacitance values are shown in Fig. 6. Since the layout of the structured separator (aka: the gate) is asymmetric upon interchanging it with respect to the position of the auxiliary electrode, one might expect respective changes in the cell's capacitance, as well. In Fig.6a-b we show the CV traces for two cases: one where the n-side of the gate faces the auxiliary electrode and the other, where the p-side of the gate faces the auxiliary electrode. The corresponding cell's capacitance values are presented in Fig. 6c for modified Cu electrodes. As observed from Fig. 6c, the cell's capacitance substantially increases upon replacing the bare separator with the diode-like gate. Additionally, capacitance increase is achieved when the p-side faces the auxiliary electrode. .

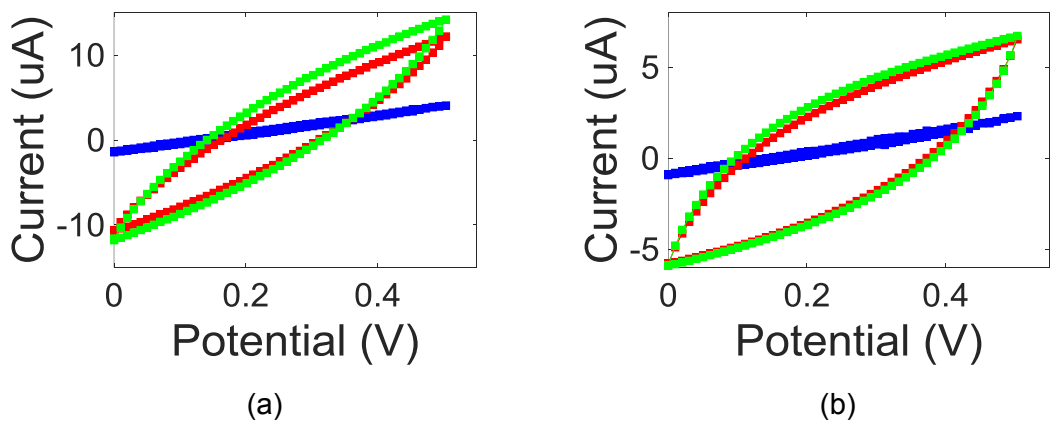

(a)              (b)

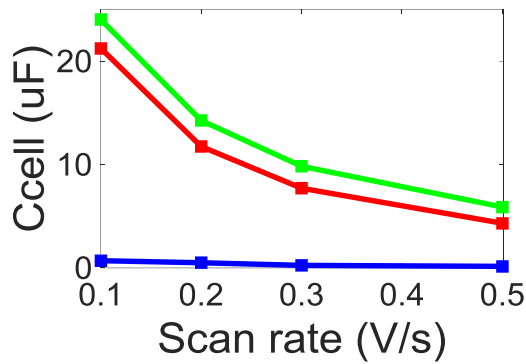

(c)

Fig. 6. Cyclic Voltammetry (CV) with two modified copper electrodes and a gate. Blue curve: bare separator; red curve: p-n structured gate where the n-side is facing the auxiliary electrode; green curve p-n structured gate where the p-side interface the auxiliary electrode. (a) At scan rate of 0.5 V/s and (b) at scan rate of 0.1 V/s. (c) Cell capacitance as a function of scan rate

Stability tests shown in Fig. 7 for 100 CV cycles. The scan rate is relatively small (0.01 V/s), hence the enhanced capacitance compared to Fig. 6c. Stability was reached after the ca 10[th] cycle. Thereafter, both configurations appear stable and varied within 2% and 4%, respectively, when either the p-side, or the n-side was facing the auxiliary electrode.

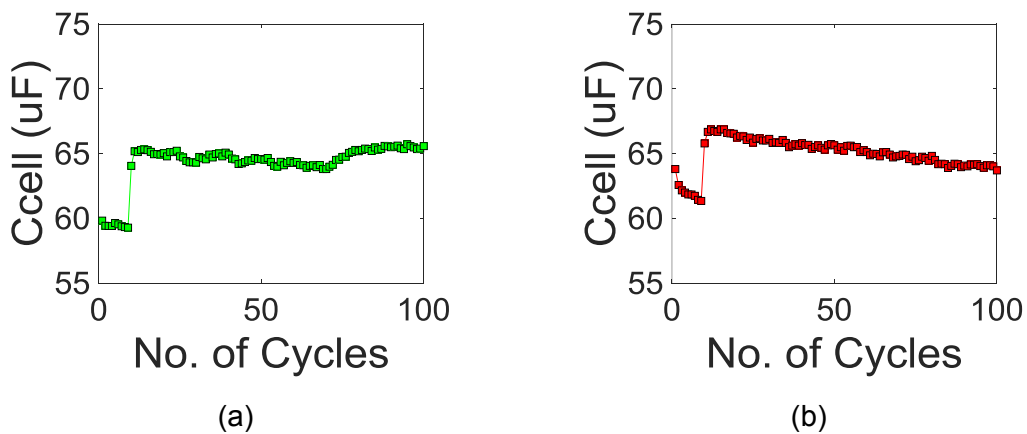

(a)  (b)

Fig. 7. Stability measurements with modified Cu electrodes. In (a) the p-side was facing the auxiliary electrode whereas in (b) the n-side was facing the auxiliary electrode. The scan rate was 0.01 V/s for both cases.

EIS measurements were performed from 50 kHz to 50 mHz with a 10 mV ac perturbation signal. In Fig. 8 we compare two cases of gate layout with respect to the auxiliary electrode. The knee frequency was 19 Hz (red) and 14 Hz (green) when the auxiliary electrode faced the n-side and the p-side, respectively. The equivalent series resistance (ESR) for ionic liquid based cells is quite high and is of the order of tens of kΩ. The smallest ESR value is noted when the p-side is facing the auxiliary electrode.

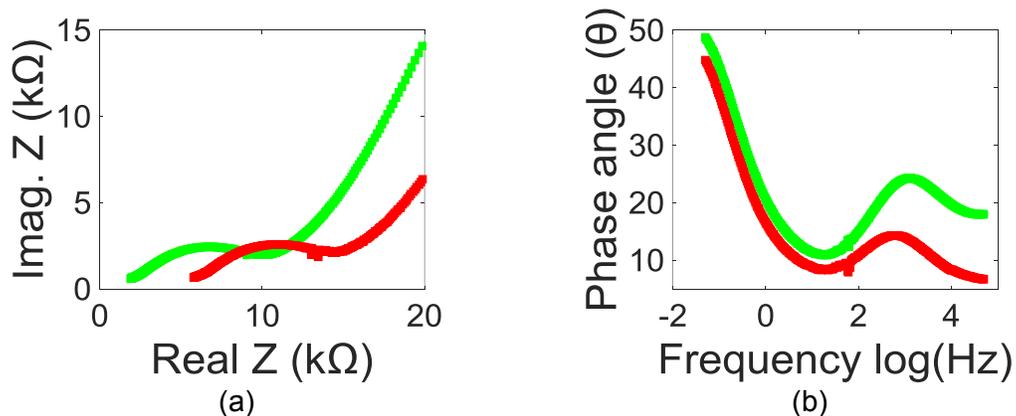

Fig. 8. (a) Nyquist and (b) Bode plots for modified Cu electrodes. Red curve: n-side is facing the auxiliary electrode. Green curve: the p-side faces the auxiliary electrode.

**IV. Conclusions:**

One can increase the capacitance of super-capacitors by incorporating a diode-like structure (called: gates) within them. The layout of the diode with respect to the auxiliary electrode affects the overall cell's capacitance. Such concept of gate within suprcapacitors (an inner gate concept), or even an outer gate structure [23] is general and can be incorporated into many cells' design. Further optimization of the gate materials, e.g., reduced graphene oxide, asymmetric cell with two types of conductive polymers and transitional metals oxides are few of many exciting possibilities.

**Declaration of interest:** We have no declaration of interest.

**Funding sources:** We have no funding sources to report.

**References**

1. R. Kötz, and M. Carlen, Principles and applications of electrochemical capacitors, Electrochimica Acta, 45 (2000) 2483.

2. John R. Miller and Patrice Simon, Electrochemical capacitors for energy management, Science, 321 (2008) 651.

3. Zhong-Shuai Wu, Guangmin Zhou, Li-Chang Yin, Wencai Ren, Feng Li and Hui-Ming Cheng, Graphene/metal oxide composite electrode materials for energy storage, Nano Energy, 1 (2012) 107.

4. Roberto Rojas-Cessa, Haim Grebel, Zhengqi Jiang, Camila Fukuda, Henrique Pita, Tazima S. Chowdhury, Ziqian Dong, Yu Wan, Integration of alternative energy sources into digital micro-grids, Environmental Progress & Sustainable Energy, 37 DOI 10.1002/ep (2018). https://doi.org/10.1002/ep.12725

5. Alberto Varzi, Corina Täubert, Margret Wohlfahrt-Mehrens, Martin Kreis, and Walter Schütz, Study of multi-walled carbon nanotubes for lithium-ion battery electrodes, Journal of Power Sources, 196 (2011) 3303

6. R.R. Nair, H.A. Wu, P.N. Jayaram, I.V. Grigorieva, A.K. Geim, Unimpeded permeation of water through helium-leak-tight graphene based membranes, Science 335 (2012) 442.

7. Y.M. Shulga, S.A. Baskakov, V.A. Smirnov, N.Y. Shulga, K.G. Belay, G.L. Gutsev, Graphene oxide films as separators of polyaniline-based supercapacitors, Journal of Power Sources 245 (2014) 33.

8. Amrita Banerjee and Haim Grebel, On the stopping potential of ionic currents, Electrochemistry Communications, 12 (2010) 274.

9. Joel Grebel, Amrita Banerjee and Haim Grebel, Towards bi-carrier ion transistor: DC and optically induced effects in electrically controlled electrochemical cell, Electrochimica Acta, 95 (2013) 308.

10. H. Grebel and A. Patel, Electrochemical cells with intermediate capacitor elements, Chemical Physics Letters, 640 (2015) 36.

11. Y. Zhang and H. Grebel, Controlling ionic currents with transistor-like structure, ECS Transactions. 2 (2007) 1.

12. S. Sreevatsa and H. Grebel, Graphene as a permeable ionic barrier, ECS Transactions, 19 (2009) 259.

13. E. Frackowiak, and F. Beguin, Electrochemical storage of energy in carbon nanotubes and nanostructured carbons, Carbon, 40 (2002) 1775.

14. M. Mastragostino and C. Arbizzani, Polymer-based supercapacitors, Journal of Power Sources, 97 (2001) 812.

15. I. H. Kim and K. B. Kim, Ruthenium oxide thin film electrodes for supercapacitors, Electrochemical and Solid State Letters, 4 (2001) A62.

16. (a) SM Mirza and H Grebel, Crisscrossed and co-aligned single-wall carbon based films, Applied Physics Letters, 91 (2007) 183102. (b) SM Mirza, H Grebel, Thermoelectric properties of aligned carbon nanotubes, Applied Physics Letters 92 (20), 203116 (2008).

17. M. J. O'Connell, P. Boul, L. M. Ericson, C. Huffman, Y. Wang, E. Haroz, C. Kuper, J. Tour, K. D. A. and R. E. Smalley, Chem. Phys. Lett., Reversible water solubilization of single wall carbon-nanotubes by polymer wrapping, 342 (2001) 265.

18. M. Shim, A. Javey, N. W. S. Kam and H. Dai, J. Am. Chem. Soc., Polymer Functionalization for Air-Stable n-Type Carbon Nanotube Field-Effect Transistors, 123 (2001) 11512.

19. A. L. M. Reddy, M. M. Shaijumon, S. R. Gowda, and P. M. Ajayan, Multisegmented Au-$MnO_2$/Carbon Nanotube Hybrid Coaxial Arrays for High-Power Supercapacitor Applications, J. Phys. Chem., 114 (2009) 658.

20. M. Kaempgen, C. K. Chan, J. Ma, Y. Cui, and G. Gruner, Nano Letts., Printable thin film supercapacitors using single-walled carbon nanotubes, 9 (2009) 1872.

21. C. Wan, L. Yuan, and H. Shen, Int. J. Electrochem. Sci., Effects of Electrode Mass-loading on the Electrochemical Properties of Porous $MnO_2$ for Electrochemical Supercapacitor, 9 (2014) 1.

22. Meryl D. Stoller and Rodney S. Ruoff, Energy Environ. Sci., Best practice methods for determining an electrode material's performance for ultracapacitors, 3 (2010) 1294.

23. H. Grebel, SN Appl. Sci., Capacitor-within-capacitor, 1 (2019) 48. https://doi.org/10.1007/s42452-018-0058-z.